# Wind-induced cross-strait sea level variability in the Strait of Gibraltar from coastal altimetry and in-situ measurements

J. Gómez-Enri[a*], C.J. González[a], M. Passaro[b], S. Vignudelli[c], O. Álvarez[a], P. Cipollini[d], R. Mañanes[a], M. Bruno[a], M. P. López-Carmona[e], A. Izquierdo[a].

[a] Applied Physics Department (University of Cadiz). Puerto Real (Cadiz), 11130, Spain.
[b] Deutsches Geodätisches Forschungsinstitut der Technischen Universität München (DGFI-TUM), Munich, Germany.
[c] CNR Institute of Biophysics (CNR-IBF), Pisa, 56024, Italy.
[d] Telespazio Vega for ESA Climate Office, Harwell, United Kingdom.
[e] Spanish Meteorological Agency (AEMET), Rota, Spain.

* Corresponding author: jesus.gomez@uca.es

**Abstract**

Coastal altimetry products are becoming available and extensively validated. Their accuracy has been assessed in many coastal zones around the world and are ready for exploitation near the shore. This opens a variety of applications of the sea level data obtained from the specific reprocessing of radar altimeter signals in the coastal strip. In this work, we retracked altimeter waveforms of the European Space Agency satellites: ERS-2 RA and Envisat RA-2 from descending track (#0360) over the eastern side of the Strait of Gibraltar using the Adaptive Leading Edge Sub-waveform (ALES) retracker. We estimated along-track Sea Level Anomaly (AT_SLA) profiles (RA-2) at high posting rate (18 Hz) using improved range and geophysical corrections. Tides were removed with a global model (DTU10) that displays a good performance in the study area: the mean root square sum (*RSS*) of the main constituents obtained with *DTU10* and 11 tide gauge stations was 4.3 cm in agreement with the *RSS* using a high-resolution local tidal hydrodynamic model (*UCA2.5D*) (4.2 cm). We also estimated a local mean sea surface reprocessing ERS-2/Envisat waveforms (track #0360) with ALES. The use of this local model gave more realistic AT_SLA than the values obtained with the global model DTU15MSS. Finally, the along-track Absolute Dynamic Topography (AT_ADT) was estimated using a local Mean Dynamic Topography obtained with the local hydrodynamic model UCA2.5D. We analysed the cross-strait variability of the sea level difference between the African / Spanish coasts along the selected track segment. This was compared to the sea level cross-strait difference from the records of two tide gauges located in the African (Ceuta) and Spanish (Tarifa) coasts. The sea level differences from altimetry and tide gauges were linked to the zonal component of the wind. We found a positive and significant (>95% c.l.) correlation between easterlies / westerlies and positive / negative cross-strait sea level differences between the southern and northern coasts of the Strait in both datasets (altimetry: r = 0.54 and in-situ: r = 0.82).

# 1. Introduction

Sea level is one of the Essential Climate Variables (ECV) listed in the Global Climate Observing System inventory (GCOS, 2016). The European Space Agency (ESA) launched in 2010 the Climate Change Initiative (CCI), aiming at providing the most accurate and homogeneous time series of some ECVs for climate studies, including altimeter-derived sea level records, such as gridded monthly maps of Sea Level Anomaly (SLA) (Quartly et al., 2017; Legeais et al., 2018) (http://www.esa-sealevel-cci.org/). Despite all the benefits of using gridded altimetry products to many applications (e.g., scientific, social, commercial), they are far from being useful in coastal zones due to poor spatio-temporal resolution of the products. This has been partially solved thanks to the efforts made in the last decade by the coastal altimetry community (http://www.coastalt.eu/community). Basically, the success is based on the generation of accurate along-track altimetry products near the shore (see Vignudelli et al. (2011); Cipollini et al. (2017); and references therein).

In recent years, many works are found in the literature focused on the generation and validation of improved along-track 'coastal' datasets from past to present altimetry missions. All of them are based on the analysis of improved geophysical corrections to the altimeter *Range* (Brown, 2010; Carrère and Lyard, 2003; Handoko et al., 2017; among others), and/or more dedicated retracking processing which take into account the shape of the radar waveforms near the coast (basically, due to land or calm water contamination) (Gommenginger et al., 2011; Passaro et al. 2014; Peng and Deng, 2018; Röscher et al., 2017). A summary table with all the datasets available to date for coastal altimetry can be found in Cipollini et al. (2017). There is no operational coastal altimetry product available yet, however, the experimental data sets that are already validated can now support the investigation of specific ocean processes occurring in regions close to land. The available experimental datasets offering retracked Ranges are the Adaptive Leading Edge Sub-waveform ALES (from Physical Oceanography Distributed Active Archive Center-PODAAC and from Open Altimeter Database-OpenADB (https://openadb.dgfi.tum.de/en/data_access/; Passaro et al., 2014) and Prototype Innovant de Systèm de Traitement pour les Applications Côtières et l'Hydrologie-PISTACH (CNES) (https://www.aviso.altimetry.fr/en/data/products/sea-surface-height-products/global/coastal-and-hydrological-products.html; Mercier et al., 2010). In a recent work, Xi-Yu et al. (2018) used a parameter available in the Jason-2 Geophysical Data Record and considered a coastal band of up to 70 km in their analysis offshore Hong-Kong. Notably, the work was an independent study that compared ALES data against data from the PISTACH coastal retracker and rated the first as the one giving the best results, while pointing out that a careful outlier analysis was needed.

Despite all these efforts, little is still done in the use of along-track high-resolution products (from 18 Hz to 80 Hz, i.e. along-track distance between two consecutive measurements from ~350 m to ~85 m) to oceanographic applications in the coastal strip. Their exploitation is crucial for many applications (infrastructure designs, coastal zone protection, and improvements in ship route security, among others) and to characterise the different processes (coastal sea level change, storm surges, coastal currents and fronts, etc.) observed in the coast. Accurate coastal altimetry products based on the reprocessing of along-track data are important for coastal observing systems (monitoring) and to re-analyse previous datasets. Recently, Dong et al. (2018) identified tidal mixing fronts using Jason-2 20-Hz along-track SLA data over Georges Bank (Northwestern Atlantic Ocean). Han et al. (2012) analysed 20-Hz and 1-Hz Jason-2 data to study Hurricane Igor storm surge off Newfoundland (Canada). At seasonal time scales, Passaro et al. (2015) determined the annual cycle of the sea level in the Baltic Sea/North Sea transition zone and Passaro et al. (2016) observed the seasonalities and trends of internal seas in the Indonesian Archipelago, using the ALES 18-Hz data set from Envisat RA-2. Here we propose to exploit the information in along-track altimetry in our study area, the Strait of Gibraltar. The product could be used to better understand and monitor the local oceanographic processes taking place in such a complex environment and to estimate transports, and therefore water exchange, contributing to 1) closing the water balance over the Mediterranean basin; and 2) monitoring changes in the Mediterranean outflow under climate change conditions.

The main contribution of this work is twofold: (1) analysis of the oceanographic content of coastal altimetry products at high-spatial resolution along track (~350 m between two consecutive 18-Hz sea level measurements) in the Strait of Gibraltar; and (2) exploitation of these products for a better understanding of the oceanographic processes in the study area. To do this, we estimated along-track Absolute Dynamic Topography (ADT) profiles obtained from the European Space Agency (ESA) Envisat RA-2 descending track #0360 (Figure 1) using the coastal-dedicated ALES retracker and accurate range and geophysical corrections. We analysed their spatio-temporal variability by estimating the sea level difference between the southern and northern sectors of the track segment and its relationship with the wind regime. The present paper is a step forward in the use of accurate altimeter products to extract relevant information of oceanographic processes close to coastal zones, such as the Strait of Gibraltar.

The paper is organised with Section 2 describing the study area (Strait of Gibraltar). Section 3 gives details of the datasets and method used to estimate the Absolute Dynamic Topography. The results are presented in Section 4 and discussed in Section 5. The paper ends with the concluding remarks in Section 6.

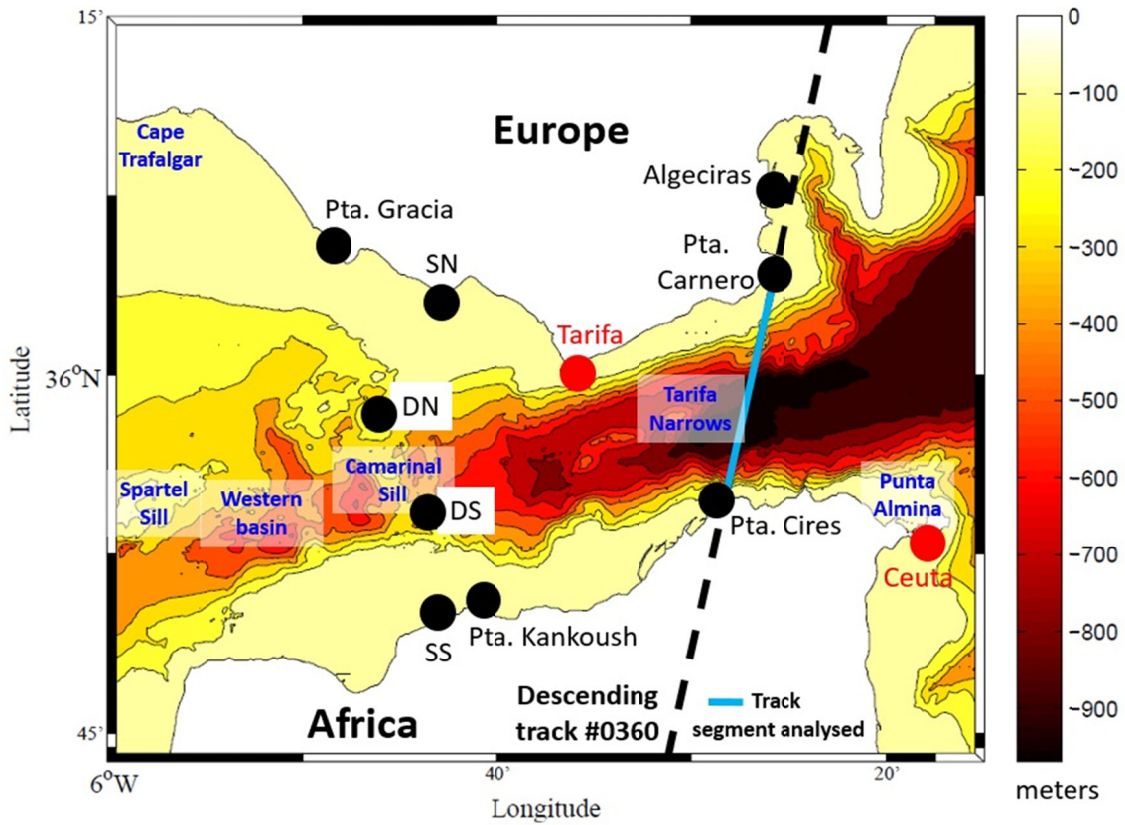

Figure 1. Study area: The Strait of Gibraltar between Europe and Africa. Colour scale indicates the bathymetry (in meters). Also shown the location of ERS-2 / Envisat descending track #0360 (the track segment analysed is highlighted in blue), the main topographic features mentioned in the text, and the location of the tide gauge and bottom pressure measurement sites used for validation of the tidal models. Red circles indicate the tide gauges used in the analysis of the cross-strait variability.

## 2. Study area: The Strait of Gibraltar

The Strait of Gibraltar is the choke point connecting the Atlantic Ocean and the Mediterranean Sea. Its bathymetry is shallower in the western part featuring the Spartel sill (depth 300 m) and the Camarinal sill (depth 280 m) separated by the Western basin. The Strait then deepens to 1100 m to the East in the Tarifa Narrows (Figure 1).

Due to the presence of less-salty, warmer Atlantic waters in the upper water masses, and saltier, cooler Mediterranean waters underneath, there is a strong density-stratification in the water column within the Strait. This leads to an inverse estuarine circulation with a two-layer exchange with a net resulting eastward flow needed to close the Mediterranean Sea water and salt balance. Lacombe and Richez (1982) characterised the hydrodynamic regime of the Strait distinguishing three main scales of variability:

(i) long period: related to seasonal, interannual, and longer period fluctuations (e.g. Garrett et al., 1990; Candela et al., 1989; Brandt et al., 2004);

(ii) subinertial: ranging from few days to few months, it is related to the exchange flows modulation caused by meteorological forcing (Candela et al., 1989). Winds are the most important meteorological forcing. Their regime is zonal with alternating easterlies-westerlies, reaching mean speeds up to 20 m s$^{-1}$. The alternated winds affect the along and cross-strait dynamics (Stanichny et al., 2005);

(iii) tidal, which represents the most energetic process in the Strait shows a predominant semidiurnal character but with a significant diurnal contribution. Tidal current velocities are strong enough to periodically reverse the flow in both layers near Camarinal Sill (Candela et al., 1990).

The Mediterranean Water outflowing the Strait of Gibraltar makes a contribution to the relatively high salinity of subsurface waters of the Nordic Seas playing a role in the deep water convection processes and, on the other side, the net flow through the Strait of Gibraltar must close the water budget over the Mediterranean basin (Candela, 2001). This is why monitoring exchange flows through the Strait of Gibraltar has been proposed as a key action for climate studies of the Mediterranean and the Global Circulation (Candela, 2001). The sea level in the Strait of Gibraltar will react to changes in the flow. Therefore, monitoring the sea level signal will provide us information on exchange flows through the Strait of Gibraltar (Hughes et al., 2015).

## 3. Datasets and methods

In this work we used accurate coastal altimetry products from the European Space Agency Satellites ERS-2 RA and Envisat RA-2 (subsection 3.1) combined with tidal models and mean sea surface models presented in subsections 3.2 and 3.3, respectively. The experimental data (tide gauge, bathymetry and wind velocity) are described in subsection 3.4. Finally, the methodology used to estimate the along-track Absolute Dynamic Topography is explained in subsection 3.5.

### 3.1 ERS-2 RA and Envisat RA-2

Along-track sea surface heights and sea level anomalies (*AT_SLA*, henceforth) 18 Hz were retrieved from the track #0360 of ERS-2 RA (cycle 1 to cycle 85, from May 1995 to June 2003) and Envisat RA-2 Phase E2 (cycle 6 to cycle 93, from May 2002 to September 2010) during the 35-day repetitive orbit of the missions. ERS-2 / Envisat RA-2 track crossed the eastern side of the Strait at 11:15 / 10:46 UTC times, respectively (Figure 1). We followed the methods explained in Gómez-Enri et al. (2016) to obtain *AT_SLA*. The altimetric datasets and ancillary corrections used and the methodology are described in full detail in the above mentioned paper, and are briefly recalled here. The *AT_SLA* was determined using the following sources of information and fields:

- Sensor Geophysical Data Records (SGDR) official product: *orbit*, *ionospheric*, *dry / wet tropospheric*, *solid earth tide and pole tide* corrections (ESA, 2007).
- Adaptive Leading Edge Subwaveform (ALES) retracker: *range*, *sigma0* and *significant wave height* (*SWH*) (Passaro et al., 2014).
- Danmarks Tekniske Universitet (DTU): *mean sea surface* (*DTU15MSS*, Andersen et al., 2016) and *tidal model* (*DTU10*, Cheng and Andersen, 2011).

Some of the corrections available in the SGDR were at a 1-Hz along-track spatial resolution, so they had to be linearly interpolated to 18 Hz prior to application to the high-rate data. *sigma0* and *SWH* from ALES were used to re-compute the *Sea State Bias* (*SSB*) correction. Gómez-Enri et al. (2016) validated time series of *AT_SLA* for the same track segment within the Strait using the same products (with the exception of the mean sea surface as they used an older version) against *in-situ* tide gauge data. They found about a 20% of improvement with respect to the *AT_SLA* computed with the *range* and *SSB* available in the SGDR official product. This was a promising result in terms of accuracy of altimetry data in the study area.

*AT_SLA* profiles (18 Hz) were estimated following Eq. (1):

$$AT\_SLA = Orbit - Range - Range\ Corrections - Geophysical\ Corrections - MSS$$

(1)

where *Orbit*, is the distance between the satellite's orbit and the WGS84 reference ellipsoid. *Range*, is the retracked range from ALES. *Range Corrections* are: *ionospheric*, *dry / wet tropospheric*, *SSB*. The set of *Geophysical Corrections* are: *solid earth, geocentric pole* and *total geocentric* ocean tides.

**3.2 Tidal models**

As mentioned, coastal altimetry products with a good level of accuracy and high spatial resolution along track (~350 m) are becoming available thanks to a number of free datasets (see Introduction section). Efforts are now needed in order to extract, assess and exploit the oceanographic content of the coastal altimetry products, especially in challenging zones such as the Strait of Gibraltar. The use of global tidal models in coastal zones to de-tide the satellite-derived sea level oscillations might be one of the main sources of noise in the sea level estimates and as such should be verified with care. In this paper, we assessed the outputs of two tidal models using in-situ measurements: (i) The global tidal model *DTU10* (Cheng and Andersen, 2011); and (ii) The local model *UCA2.5D* (Izquierdo et al., 2001).

*3.2.1 Global tidal model DTU10*

The model *DTU10* is an updated version of *FES2004* (Lyard and Lefèvre, 2006) ocean tide. Its resolution is 0.125º x 0.125º worldwide. The tidal elevations (as the sum of two terms: *Ocean Tide* and *Loading Tide*) were extracted from the Danmarks

Tekniske Universitet ftp server: (ftp://ftp.space.dtu.dk/pub/DTU10/). *DTU10* was interpolated to the along-track positions of the track segment analysed.

*3.2.2 Local model: UCA2.5D*

The two-dimensional (depth-averaged), two-layer, finite-difference, hydrodynamic model *UCA2.5D* (Izquierdo et al., 2001; Brandt et al., 2004) was developed specifically to reproduce the two-layer dynamics in the highly density-stratified environment of the Strait of Gibraltar. The model outputs are the depth-averaged horizontal currents in the upper and lower layers, as well as their respective top-limit heights (i.e., free-surface and interface depth). The curvilinear model grid extends from the easternmost area of the Gulf of Cadiz to the western half of the Alboran Sea, with a variable horizontal spatial resolution ranging from ~1 km in the eastern and western boundaries of the domain to ~125 m within the Strait of Gibraltar. The model currently runs in pre-operational mode and is forced at the open boundaries by the main semidiurnal ($M_2$, $S_2$) and diurnal ($K_1$, $O_1$) tidal constituents, as well as by surface wind and Mean Sea Level Pressure (MSLP) fields (Izquierdo et al., 2016) provided by the meteorological model MM5 (www2.mmm.ucar.edu). *UCA2.5D* has been extensively applied and validated in the Strait of Gibraltar, showing in general a good agreement with the observed hydrodynamics (Sein et al., 1998; Izquierdo et al., 2001; Brandt et al., 2004; Izquierdo et al., 2016).

**3.3 Mean Sea Surface models**

A mean sea surface (MSS) profile is subtracted to the along-track corrected sea surface height in order to get the anomalies (SLA). We used two sources of along-track

MSS: (i) Global Mean Sea Surface: *DTU15MSS*; and (ii) Along-track Local Mean Sea Surface: *AT_Local_MSS*.

*3.3.1 Global Mean Sea Surface: DTU15MSS*

This mean sea surface model is an updated version of the previous *DTU10MSS* (Andersen and Knudsen, 2009; Andersen, 2010). The new version includes, globally, 4 years of CryoSat-2 data from its three modes of operation: Low Rate Mode (LRM), Synthetic Aperture Radar (SAR) and SAR-Interferometric (SAR-In). Data were extracted from the Danmarks Tekniske Universitet ftp server: ftp://ftp.space.dtu.dk/pub/DTU15/1_MIN/ at 1-minute resolution grid. *DTU15MSS* was interpolated to the along-track positions of the track segment analysed.

*3.3.2 Along-track Local Mean Sea Surface: AT_Local_MSS*

An along-track Mean Sea Surface (*AT_Local_MSS*) based on ALES data was computed by interpolating the along-track Sea Surface Height (SSH) onto nominal tracks following the procedure explained in Passaro et al. (2014). We used the descending orbit #0360 combining all the overpasses from both the ERS-2 and Envisat missions with the same set of corrections. This approach brings several beneficial effects:

- Using two missions allows the computation of the anomalies in the study area, where the variability is higher, with a local mean sea level measured at the same track locations over a longer time span than from the Envisat mission alone;

- Using the same algorithm (ALES) and the same corrections as noted by Andersen and Scharroo (2011), the use of a MSS computed using altimetry data that are retracked with different algorithms and could therefore suffer from biases w.r.t. ALES data even in the open ocean prevents biases or differences that might otherwise show up as dynamic topography.

**3.4 Experimental data sets**

*3.4.1 Tide gauge and bottom pressure data*

The validation of the global (*DTU10*) and local (*UCA2.5D*) tidal models was made using information available in the literature (García-Lafuente, 1986; Candela et al., 1990). We used the amplitude and phase of the main tidal constituents obtained from 11 *in-situ* instruments (tide gauges and bottom pressure sensors): Tarifa, SN, DN, DS, SS, Pta. Gracia, Pta. Kankoush, Pta. Carnero, Pta. Cires, Algeciras, and Ceuta (Figure 1).

The cross-strait sea level variability in the study area was also analysed using the measurements from two pressure tide gauges deployed at Ceuta and Tarifa by Puertos del Estado (http://www.puertos.es) (Figure 1). The recorded 5-minute resolution time series of water column height from 2002 to 2010 were smoothed and decimated to standard hourly values (Godin, 1972), and the resulting hourly series were subjected to harmonic analysis (Foreman and Henry, 1989), in order to obtain the amplitude and phase harmonic constants for the resolvable tidal constituents, as well as the mean sea level relative to the instrument. The difference between the original time series and the corresponding tidal prediction computed from the obtained tidal harmonics, usually known as 'residual height', is assumed to be free of any tidal contribution, and the resulting residual series from the two locations were used to obtain the sea level differences between them, in order to analyse their relationship with the wind conditions.

*3.4.2 Bathymetry*

Bathymetry has a high spatial resolution of 50 m, provided by the Spanish Navy Hydrographic Institute. It is available by request to the EMODNET Project web page: http://www.emodnet-bathymetry.eu.

*3.4.3 Wind velocity*

The zonal component of the wind velocity ($u$) used in the study was extracted from the hourly time series of 10-m height wind speed and direction from October 2002 to October 2010, recorded by the weather station deployed by the Spanish Meteorological Agency (AEMET) at Tarifa (Figure 1).

**3.5 Absolute Dynamic Topography (*ADT*)**

The sea level above geoid is commonly known as Absolute Dynamic Topography and can be obtained from the sum of two terms: along-track Sea Level Anomaly (*AT_SLA*) and along-track Mean Dynamic Topography (*AT_MDT*):

$$AT\_ADT = AT\_SLA + AT\_MDT \qquad (2)$$

*AT_SLA* was obtained using the corrections presented in subsection 3.1 (Eq. 1). *AT_MDT* was obtained from two sources of Mean Dynamic Topography: *Global / Local MDT*.

*3.5.1 Global MDT*

We used the most updated version available of the DTU Mean Dynamic Topography. *DTU15MDT* (Knudsen et al., 2016) were extracted from the ftp server: ftp://ftp.space.dtu.dk/pub/DTU15/1_MIN/ at 1-minute resolution grid. In this model the newer version of the gravity field (EIGEN-6C4) has been combined with *DTU15MSS* to obtain the global Mean Dynamic Topography.

*3.5.2 Local MDT*

The local MDT was obtained as the time mean of 1-year long sea surface height hourly output from the local tidal model: *UCA2.5D* (Izquierdo et al., 2001). Brandt et al. (2004) validated this numerical model in the study area in terms of current velocity (current-meter moorings and the inverse tidal model of Baschek et al. 2001), and sea surface elevation (bottom pressure tide gauges). In general, the authors found a good level of agreement between the numerical tidal model, the observations and the inverse tidal model of Baschek et al. (2001).

**4. Results**

**4.1 Assessment of the models used for altimeter corrections**

*4.1.1 DTU10 vs. Local Tidal model*

The performance of the global / local tidal models in the Strait is analysed in detail. The global tidal model *DTU10* was already assessed in a previous work (Gómez-Enri et al., 2016). The authors compared the main constituents derived from a tide gauge located at Tarifa harbour with the constituents provided by *DTU10* (at the tide gauge location) estimating a root square sum (*RSS*) of 4.6 cm, following Oreiro et al.

(2014). We extended this analysis by assessing the models (*DTU10* and *UCA2.5D*) estimating the *RMS* of the differences and the *RSS* of the main constituents: $M_2$ and $S_2$ (semidiurnals), $K_1$ and $O_1$ (diurnals) using information from a few tide gauges and bottom pressure instruments located within the Strait of Gibraltar (Figure 1). The results are summarized in Table 1. Overall, both models show similar results, in terms of *Total_RSS* (4.3 / 4.2 cm for *DTU10* / *UCA2.5D*, respectively); we observe the lowest *RSS* (below 2.0 cm) in a few stations: DN and Algeciras (*DTU10*), SS and Ceuta (*UCA2.5D*). In the case of *DTU10* this might be due to the fact that these stations are likely assimilated in the model. The comparison with the closest stations to the satellite track (Pta. Carnero and Pta. Cires) gives similar *RSS* in Pta. Cires for both models and slightly better results for *DTU10* in Pta. Carnero.

The analysis of the tidal models in the study area indicates that *DTU10* shows a similar level of accuracy as the local model. For this reason, we used the global model to de-tide the sea level signals to estimate the *AT_SLA*.

*4.1.2 Spatial variability of AT_SLA*

Not all the overpasses of the satellite yield usable data – some cycles are missing due to acquisition problems or other platform issues and in a few cases the data were collected with a lower radar pulse bandwidth (20 MHz or 80 MHz instead of the usual 320 MHz over ocean), which makes them not accurate enough for oceanographic investigation. The number of overpasses with valid 320-MHz data is 78. The removal of outliers was made in two steps: (i) Only the *SLA* values between [-2.5 2.5] (m) were retained; and (ii) All the absolute *SLA* values greater than 3 times the standard deviation of the mean along-track profile were considered as outliers. A five-elements running mean was applied to *AT_SLA* to remove high frequency noise (equivalent to 1.75 km).

Table. 1. *RMS* of the differences of the main constituents from the global DTU10 and local UCA2.5D tidal models with some tide gauge and bottom pressure instruments in the study area. The *RSS* and *Total_RSS* are also shown.

|  | *RMS* differences (cm) | | | | *RSS* (cm) |
|---|---|---|---|---|---|
|  | $M_2$ | $S_2$ | $K_1$ | $O_1$ |  |
| Station/Tide Gauge | DTU10 / UCA2.5D | | | | DTU10 / UCA2.5D |
| Tarifa | 4.9 / 5.6 | 1.9 / 2.0 | 0.6 / 1.2 | 1.3 / 1.7 | 5.5 / 6.3 |
| SN | 4.9 / 5.6 | 2.1 / 1.9 | 0.4 / 1.0 | 0.6 / 1.1 | 5.4 / 6.1 |
| DN | 0.8 / 4.8 | 1.0 / 2.4 | 0.3 / 0.9 | 1.0 / 0.7 | 1.7 / 5.5 |
| DS | 4.0 / 2.4 | 0.2 / 0.9 | 0.9 / 1.0 | 0.9 / 1.2 | 4.2 / 3.0 |
| SS | 5.5 / 1.3 | 2.0 / 0.3 | 1.7 / 1.0 | 1.6 / 0.8 | 6.3 / 1.9 |
| Pta. Gracia | 3.6 / 6.3 | 2.2 / 2.2 | 1.1 / 1.5 | 0.7 / 1.5 | 4.4 / 7.0 |
| Pta. Kankoush | 7.0 / 3.4 | 1.5 / 1.3 | 0.8 / 1.7 | 1.2 / 1.8 | 7.3 / 4.5 |
| Pta. Carnero | 1.0 / 2.4 | 0.9 / 0.03 | 0.5 / 0.5 | 1.3 / 0.7 | 2.0 / 2.6 |
| Pta. Cires | 4.6 / 4.6 | 1.2 / 1.3 | 0.5 / 0.8 | 1.1 / 0.6 | 4.9 / 4.9 |
| Algeciras | 0.6 / 2.8 | 0.4 / 0.2 | 0.7 / 0.3 | 0.6 / 0.7 | 1.2 / 2.9 |
| Ceuta | 3.6 / 0.8 | 1.8 / 0.4 | 0.1 / 0.4 | 0.2 / 0.3 | 4.0 / 1.0 |
|  |  |  |  |  | *Total_RSS* 4.3 / 4.2 |

As shown from the results in Table 1 the accuracy of the global tidal model *DTU10* needs to be assessed with care in the Strait of Gibraltar. This model does not include tidal constituents of longer periods than semidiurnal and diurnals (Cheng and Andersen, 2011). Two of the most important constituents of longer periods in the Strait are: $M_{sf}$ (lunisolar synodic fortnightly) and $M_m$ (lunar monthly). The harmonic analysis made to de-tide the in-situ time series at Ceuta and Tarifa stations (Figure 1), included these and others constituents. We analysed their magnitude in the vicinity of the satellite track segment. To do this, we obtained their amplitude and phase from the literature (Garcia-Lafuente et al., 1990) at the closest in-situ stations to the satellite pass: Pta. Carnero and Pta. Cires (Figure 1). The values are summarised in Table 2. The fortnightly constituent ($M_{sf}$) has a small and similar amplitude at both sides of the Strait showing a similar phase. The amplitude of the lunar monthly ($M_m$) is of the same order of magnitude at both stations, but they are in phase opposition. This comparison might indicate a small impact of not using longer period constituents to de-tide the sea level with the global model DTU10 in the Strait.

Table 2. Amplitude and phase of $M_{sf}$ and $M_m$ constituents for Pta. Carnero and Pta. Cires tide gauge stations

|  | **Pta. Carnero** | | **Pta. Cires** | |
|---|---|---|---|---|
|  | *Amplitude (cm)* | *Phase (º)* | *Amplitude (cm)* | *Phase (º)* |
| **$M_{sf}$** | 2.5 | 92 | 1.4 | 40 |
| **$M_m$** | 1.1 | 153 | 1.9 | 302 |

A visual inspection of the radargrams of the waveforms gave two main patterns in the leading edge area of the returned echoes: (i) 'stable' leading edge around the nominal tracking point; this was found in 30 out of 78 cycles; (ii) 'non-stable' undulated leading edge in the northern sector of the track segment; this was observed in 42 cycles. In 3 cycles we noted a strange behaviour in the leading edge and thermal noise areas. Finally, 3 cycles showed a short track segment. Thus, only 40% of cycles (30) were considered valid for further analysis. The radargrams of the waveform power of a rejected cycle (no. 6) and a valid cycle (no. 14) is presented in Fig. 2a and Fig. 2b, respectively. The undulated radargram is due to the known problems of the on-board tracker in keeping the signal within the analysis window in proximity of land (Gommenginger et al. 2011).

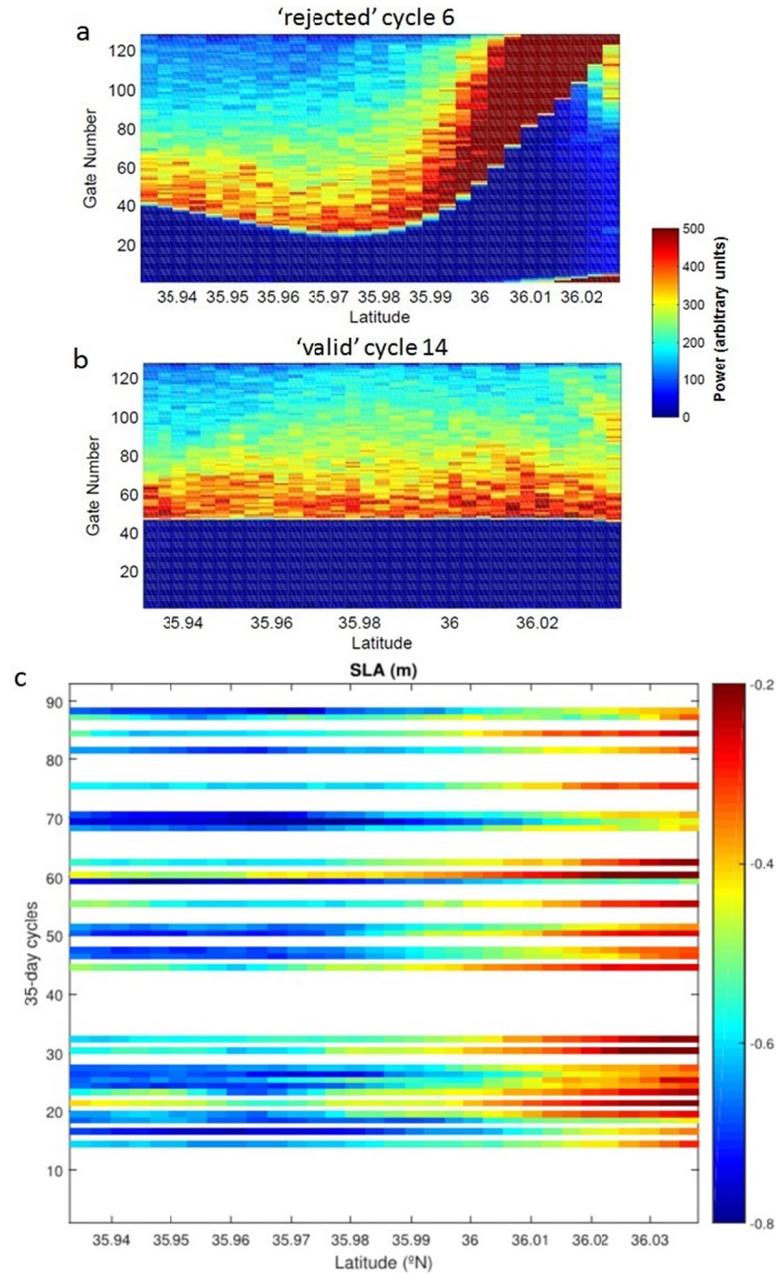

Figure 2. Radargram of the RA-2 radar waveform power for a rejected (Fig. 2.a) and a valid (Fig. 2.b) cycle. The cycle-by-cycle *AT_SLA* for valid cycles is shown in Fig. 2.c.

The cycle-by-cycle *AT_SLA* profiles after the data screening and the removal of invalid cycles due to unstable leading edges in the waveforms are shown in Fig. 2c. The sea level oscillates in the range: [-0.8 -0.2] m with a marked difference between the African (lower sea level) and the Spanish (higher sea level) coasts. This behaviour is not in agreement with previous works made in the study area. Ross et al. (2000) indicated higher sea level on the southern side of the Strait. They used data from two tide gauges located in Ceuta (African coast) and Algeciras (Spanish coast) (Figure 1). The magnitude of this difference seems to range between 20 cm and 10 cm between maximal and submaximal exchange (Bormans and Garrett, 1989). A similar result was also reported by Stanichny et al. (2005).

One of the reasons that could explain this disagreement might be in the use of a global mean sea level (*DTU15MSS*) to estimate *AT_SLA*. The model, as the other global MSS models in the literature, is derived by merging several years of data from repeated tracks with nonrepeating data from geodetic missions using sophisticated interpolation techniques. The accuracy of these models is known to be degraded close to the coast (Andersen et al., 1999; Vignudelli et al., 2006), since the quality and the amount of the altimeter data used in the models is notably lower in the last ~25 km from the coast (Passaro et al. 2014; 2015), which means that the MSS values in the coastal zone are generally extrapolated (Andersen et al., 1999). This might be particularly relevant in our study area due to the lack of accurate along track high-resolution altimeter data in the past (Gómez-Enri et al., 2016). We investigated this by comparing the *AT_SLA* profiles obtained using the global (*DTU15MSS*) and the local (*AT_MSS*) mean sea levels. Figure 3 shows the average of *AT_SLA* for over all valid cycles. Fig. 3a (*DTU15MSS*) gives a lower sea level in the African coasts (5-elements running mean applied). The ERS2 / Envisat derived mean sea surface gives an *AT_SLA* (Fig. 3b) with a marked positive cross-strait sea level difference between the African and Spanish coasts, which is in agreement with previous studies (Bormans and Garrett, 1989; Ross et al., 2000; Stanichny et al., 2005). Thus, the 'local' mean sea level (*AT_MSS*) was selected in order to estimate the along-track Absolute Dynamic Topography (*AT_ADT*).

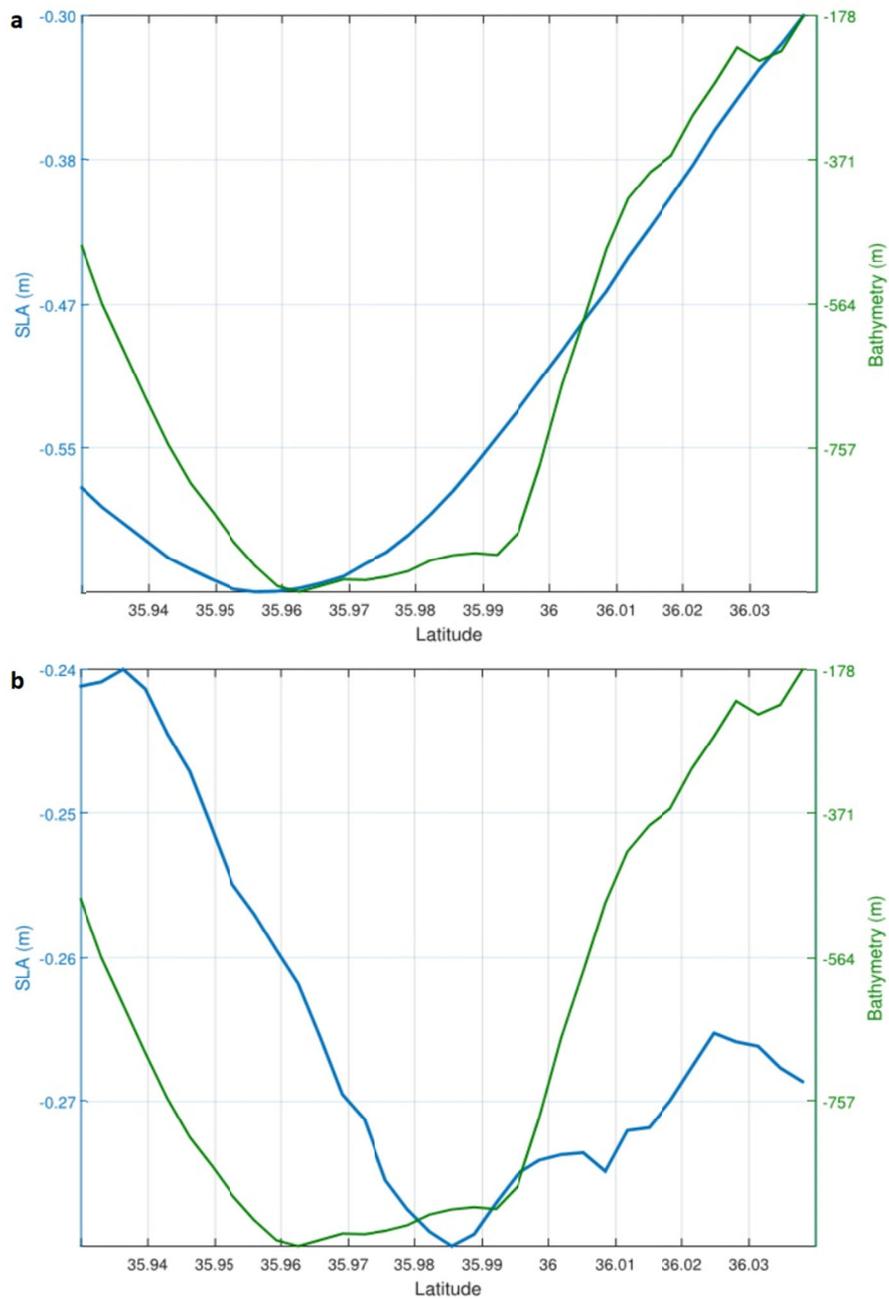

Figure 3. *AT_SLA* (5-elements running mean applied) for valid cycles using the global *DT15MSS* (Fig. 3.a) and the local *AT_MSS* (Fig. 3.b). The green line gives the bathymetry profile.

## 4.2 Cross-strait variability in the Strait of Gibraltar

*4.2.1 Along-track ADT*

As mentioned, we used two sources of *MDT* to compute the *AT_ADT* (Eq. 1). Figure 4 shows the *MDT_DTU15* (Fig. 4a) and *MDT_Local* (Fig. 4b) obtained from the local numerical model *UCA2.5D*. The magnitude of *MDT_DTU15* ranges between 4.5 - 7.5 cm, smaller than *MDT_Local*: 0 - 10 cm. The zonal variation of *MDT* (sea level decrease toward east) along the strait can be observed in both models but the local *MDT* shows a remarkable meridional cross-strait gradient. We also observe a difference in the magnitude of the mean along-strait sea level difference between the two entrances of the Strait. Thus, this difference is smaller than 2 cm for the global *MDT* but up to 6 cm for the local model. Also, the strong change in sea level pattern around Camarinal Sill shown in *MDT_Local* is not observed in *MDT_DTU15*. A possible explanation might lie in the fact that this global model does not resolve properly the baroclinic coupling between upper and lower layer flows (Brandt et al., 2004).

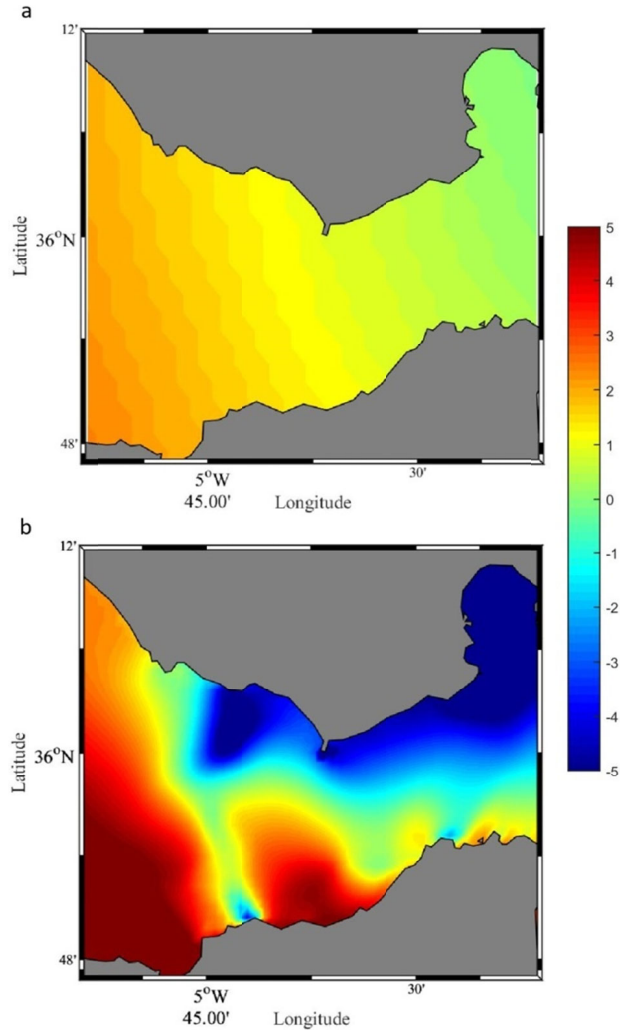

Figure 4. Mean Dynamic Topography (in cm) in the Strait of Gibraltar from the global model *MDT_DTU15* (5.a) and the local model *MDT_UCA2D* (5.b).

We recalculated along-track *ADT* (*AT_ADT*) following Eq. (2) using the *Local_MDT* (only for cycles with a 'stable' leading edge around the nominal tracking point). It can be seen a clear cross-strait sea level difference with higher sea level on the African side in most of the cycles (Fig. 5.a). Fig. 5.b gives the *AT_ADT* for a cycle (no. 51) with a positive cross-strait sea level difference. Also, inversions of the sea level differences between south and north sides are observed in some cycles. Fig. 5.c shows *AT_ADT* for cycle 50. We investigated this further by analysing its relationship with the wind regime.

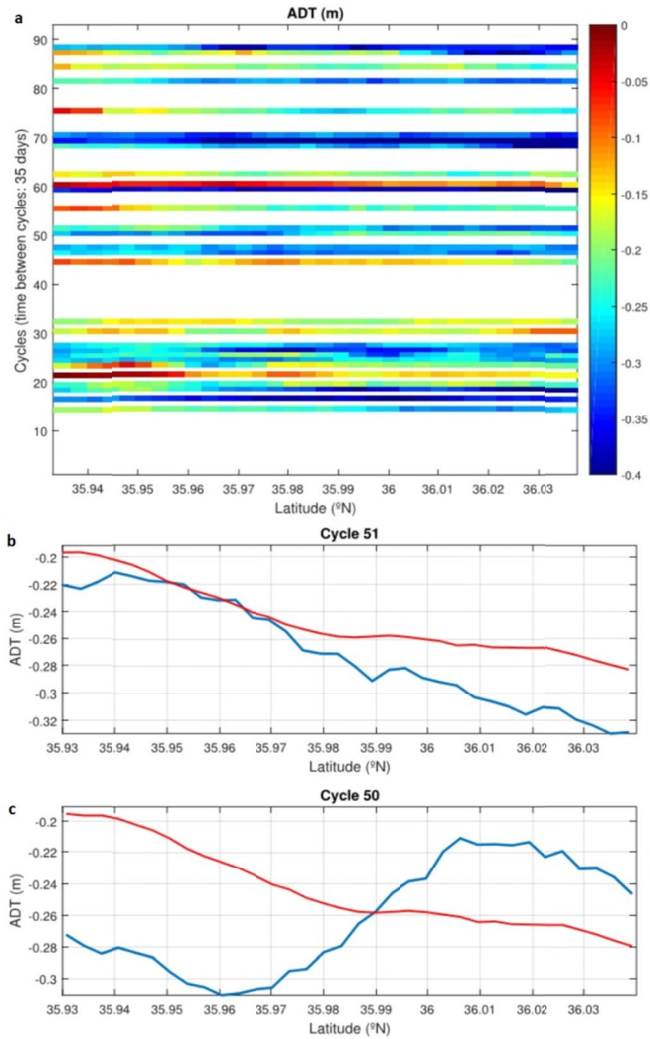

Figure 5. Cycle-by-cycle *AT_ADT* for valid cycles (Fig. 5.a). *AT_ADT* for cycle 51 (Fig. 5.b) and 50 (Fig. 5.c). The average of *AT_ADT* over all valid cycles is also shown in Fig. 5.b and 6.c (red line).

*4.2.2 Inversion of the cross-strait sea level difference due to wind regime*

We analysed the variability of the cross-strait sea level in the Strait of Gibraltar and its dependence on the wind regime. We first determine the response time of sea level to wind forcing following an empirical approach: the wind mean zonal component ($\hat{u}$) was calculated as the average $u$ value from a time interval ($N_h$) starting immediately before the pass of the satellite (about 10:46 UTC time). $N_h$ was ranging from 1 to 48 hours before that time. The time response selected was that corresponding to the averaging interval giving the maximum correlation coefficient between the calculated zonal component ($\hat{u}$) and the sea level difference between the two tide gauge stations. Figure 6 gives the correlation coefficients (p_value < 0.05) between $\hat{u}$ and the sea level gradient for increasing $N_h$. The best correlation (r = 0.812) was found for $N_h$ = 2. Easterly/westerly winds (the so-called Levante/Poniente winds, respectively) give negative/positive $\hat{u}$ component.

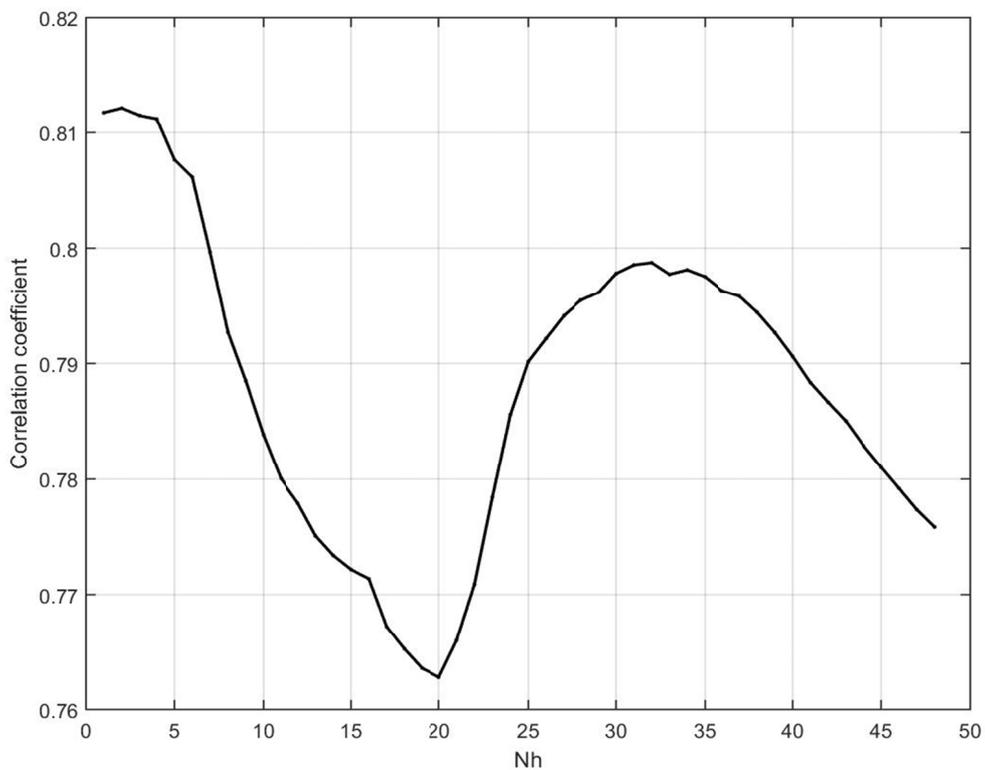

Figure 6. Correlation coefficient between the calculated zonal component ($\hat{u}$) and the sea level gradient between the two tide gauge stations (Ceuta and Tarifa) for increasing time intervals ($N_h$).

We obtained the sea level difference between the south and north sides of the Strait from altimetry and the two tide gauges located in Ceuta and Tarifa (Figure 1). The sea level difference from altimetry was obtained by estimating the cross-strait sea level slope using the starting and ending points of the linear fit made to the *AT_ADT* profiles. We found 8 out of 30 cycles with a small confidence level and were rejected in the analysis. The relationship between $\hat{u}$ and the sea level differences are shown in Figure 7 for altimeter (Fig. 7.a) and tide gauge data (Fig. 7.b). Negative differences indicate the inversion of the sea level difference between the southern and northern sectors of the Strait. This is clearly seen during easterlies in the two datasets, more evident in the tide gauge data, where the correlation coefficient is remarkably high (0.82). The inversion is due to the Ekman transport during strong easterlies. As mentioned, this is in agreement with previous observations in the study area. Stanichny et al. (2005) analysed correlation between the sea level gradient among tide gauges located in Ceuta, Algeciras and Tarifa and the zonal wind component of velocity. The authors found negative cross-strait differences during severe easterlies and a notably similar correlation coefficient (0.85).

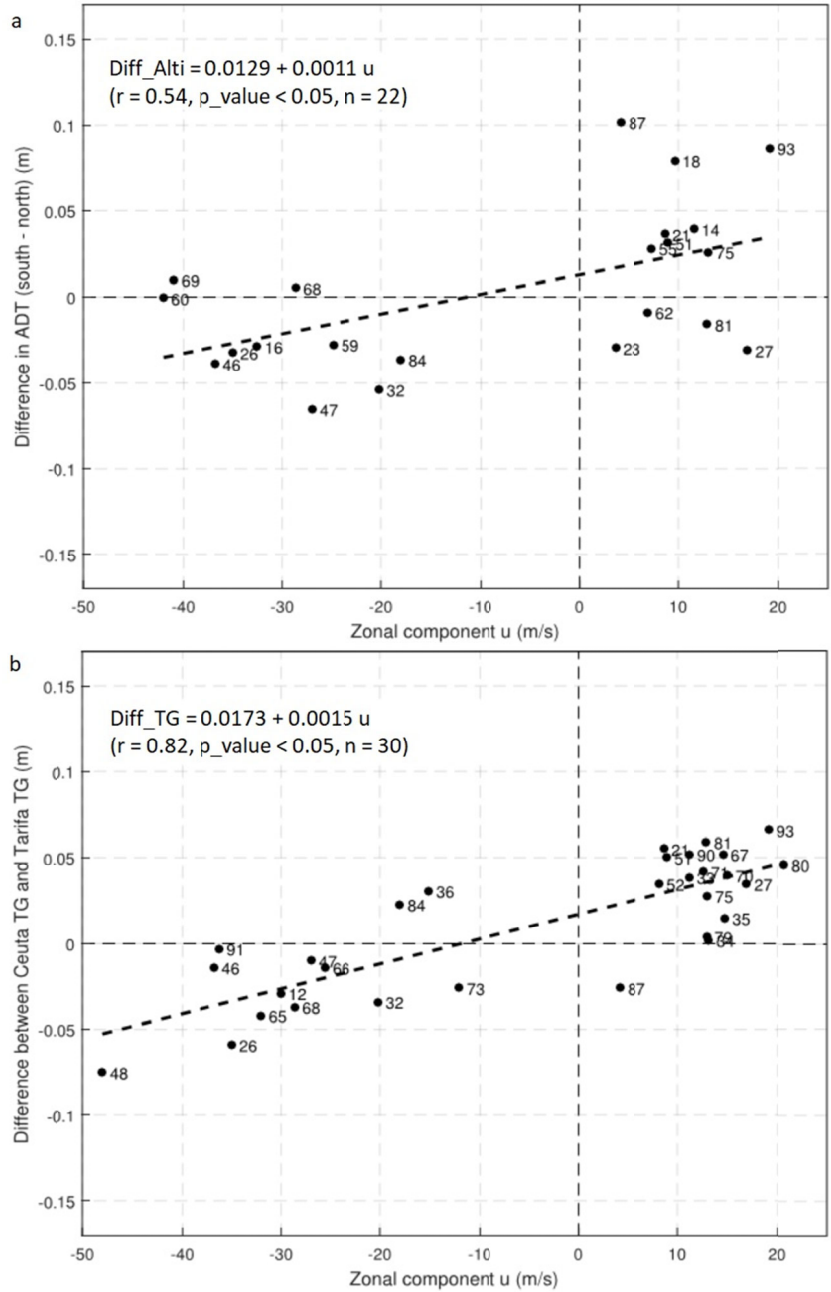

Figure 7. Cross-strait sea level differences between the African and Spanish coasts as a function of the mean zonal $\hat{u}$ component of the wind. Fig. 7.a gives the comparison using altimeter data and Fig. 7.b in-situ data. The numbers indicate the cycle number of the satellite pass.

## 5. Discussion

Our work was focused on the use of an accurate coastal sea level product from a pulse-limited altimeter (RA-2) in the Strait of Gibraltar. We have highlighted how the studies that look at spatial SLA profiles should take particular care of the MSS adopted.

We have shown (Figure 3) that even the use of a dedicated retracking procedure might not be able to provide a realistic sea level profile if the MSS used to extract the anomaly is based on standard products. There are at least two explanations for this. Firstly, the standard products are severely flagged in the coastal zone to avoid the risk of erroneous retrievals, which means that a global MSS model is probably characterised by strong data interpolation in the absence of local sea level data in areas such as the Strait of Gibraltar. Secondly, the use of different geophysical corrections in the computation of the global MSS w.r.t. the corrections used in the reprocessed SSH might leave unrealistic features when deriving the anomalies. This is for example the case of corrections with different spatial resolutions: for example, the Wet Tropospheric Correction from the radiometer rather than from an atmospheric model, or the Dynamic Atmosphere Correction forced with the operational ECMWF atmospheric model rather than with the ERA-Interim Reanalysis. These problems have been mentioned by Andersen and Scharroo (2011), and more recently by Han et al. (2017), but the Coastal Altimetry community has up to now overlooked the problem, which becomes of urgent matter nowadays, when the discipline is mature for oceanographic exploitation.

A possible future extension of this work would be the use of SAR mode altimetry from CryoSat-2 and Sentinel-3. The accuracy of current delay-Doppler missions: CryoSat-2 SIRAL and Sentinel-3A SRAL (already joined by Sentinel-3B) in terms of sea level has not been quantified in the study area. A thorough validation exercise of CryoSat-2 SIRAL data was performed by Gómez-Enri et al. (2017) in the eastern shelf of the Gulf of Cadiz (western end of the Strait). They concluded that the quality of CryoSat-2 20-Hz SLA data was comparable to conventional altimetry (Saral AltiKa 20-Hz data) in the coastal strip of the Gulf. However, the complexity of the Strait of Gibraltar precludes the use of SAR altimetry missions for scientific exploitation before performing their validation with in-situ tide gauge data. The unique orbital configuration of the CryoSat-2 mission (369-days repeated cycle with a 30-days sub-orbital cycle) would allow a geographically distributed assessment of the cross-strait variability of the sea level in the Strait and its relationship with the wind regime. In addition to this, the 27-days orbital cycle of Sentinel-3A gives two tracks (one ascending and one descending) very close to the Envisat track analysed in this work with the cross-over at Algeciras city, north of Punta Carnero (Figure 1). This convenient sampling by Sentinel-3A combined with its SAR mode should allow a better knowledge of the hydrodynamic processes in the eastern side of the Strait of Gibraltar. For both CryoSat-2 and Sentinel-3, the use of a local mean sea surface to obtain the anomalies is needed.

## 6. Conclusions

The cross-strait variability in the eastern side of the Strait of Gibraltar and its relation to the wind regime has been analysed. We estimated the sea level differences between the southern and northern sector of the Strait and analysed their variability as a function of the zonal component of the wind. To do this, we estimated the absolute dynamic topography from the sea level anomalies obtained using along-track SLA (Envisat RA-2 descending track #0360) based on ALES reprocessing and improved geophysical corrections. The sea level differences and its dependence with the wind regime were compared with the outputs of two tide gauges located at both sides of the Strait. The main conclusions are summarised as follows.

The global mean sea level *DTU15MSS* does not take into account some of the particularities of such a complicated zone as the Strait of Gibraltar. Our results show that its use to compute the sea level anomalies might hide some of the sea level variability, and hence complicate their oceanographic interpretation. We have demonstrated that a local mean sea level (*AT_Local_MSS*) based on ALES reprocessing of ERS2/Envisat descending track #0360 along-track sea surface heights, gives a more realistic cross-strait variability in the Strait improving the analysis of the hydrodynamic processes in the area. The global tidal model *DTU10* shows a good performance in the Strait of Gibraltar to de-tide altimetric records. The mean *RSS* of the main constituents obtained with *DTU10* and 11 stations is 4.3 cm very similar to the *RSS* using a local tidal model (*UCA2.5D*) (4.2 cm).

The cross-strait variability obtained between the southern (Pta. Cires) and northern (Pta. Carnero) eastern zone of the Strait of Gibraltar is highly dependent on the wind regime. The analysis of the along-track absolute dynamic topography showed a positive correlation with the zonal component of the wind. The difference between the sea level in the southern and northern segments of the altimeter track gives positive / negative differences under westerlies / easterlies conditions. The inversion (negative) of that difference is related to severe easterlies as a result of the Ekman transport. This contributes to the modulation of the water exchange through the Strait of Gibraltar, weakening the Atlantic water inflow toward the Mediterranean Sea.

Coastal altimeter data are ready for exploitation in the coastal zone. Data accuracy needs to be continuously assessed near the shore, especially the products obtained from present altimeter (conventional and delay-Doppler) missions. The Strait of Gibraltar, the unique connection between the Mediterranean Sea and the Atlantic Ocean, plays an important role in their water exchange and its complexity has been thoroughly investigated for many years using ground-truth instruments and its hydrodynamic processes have been modelled. We have demonstrated here that satellite altimetry gives accurate sea level measurements in the Strait, and hence helps to a better knowledge of its hydrodynamic processes. We have also shown that, at least regionally, coastal altimetry can be used to significantly improve the knowledge of the MSS in the coastal zone. There is therefore a strong need to perform an impact assessment looking at the differences between current global MSS models and along-track MSS computed using reprocessed dataset such as ALES in other regions. High quality satellite-based altimeter with finer along-track spatial resolutions from SAR-mode missions such as Sentinel-3A/B should be incorporated to the experimental datasets available in the area.

**ACKNOWLEDGMENTS**

The authors are thankful to the Spanish Instituto Hidrográfico de la Marina (IHM) for the bathymetric dataset, and the Spanish Puertos del Estado for the tide gauge data at Tarifa and Ceuta. Special thanks to Ole B. Baltazar (Denmark Technological University) for his comments on the global tidal model DTU10. This work was partially funded by the OCASO-Interreg POCTEP project.